\documentclass[11pt,a4paper]{article}
\usepackage[pdftex]{graphicx}
\usepackage[english]{babel}
\usepackage[utf8]{inputenc}
\usepackage{times}
\usepackage{wasysym}
\usepackage{cite}
\usepackage[square,numbers]{natbib}

\def\beqn{\begin{eqnarray}}
\def\eeqn{\end{eqnarray}}

\def\beq{\begin{equation}}
\def\eeq{\end{equation}}

\newcommand{\mc}{\multicolumn}

\def\eex#1{$ \times 10^{#1}$}
\def\eexeq#1{ \times 10^{#1}}

\def\lesssimA#1#2{\mathrel{\vcenter{\offinterlineskip%
    \ialign{\hfil##\hfil\cr$#1<$\cr$#1\sim$\cr}%
}}}
\def\lesssim{\mathpalette\lesssimA{}}

\def\ip1j{^{i+1,j}}
\def\im1j{^{i-1,j}}

\def\ijp1{^{i,j+1}}
\def\ijm1{^{i,j-1}}
\def\ijph{^{i,j+1/2}}

\def\enhk{\epsilon_{k}}

\def\meaneq#1{\langle #1 \rangle} 
 
\def\D{\Delta}

\begin{document}

\pagenumbering{arabic}
\pagestyle{plain} 


\noindent  {\bf \Large Adjacent Sink Strengths Used in Multiscale Kinetic Rate Equation Simulations of Defects and Impurities in Solids}\\

\noindent Tommy Ahlgren,\\
\noindent University of Helsinki, P.O. Box 43, FI-00014 Helsinki\\  
\noindent March 3, 2026
\tableofcontents

\section{Abstract}
Kinetic Rate Equation (kRE) modeling is widely used to simulate defect and impurity evolution in solids over experimentally relevant time and length scales.
However, conventional kRE formulations include only random-position sink strengths, which adequately describe trapping of
defects created at random lattice sites but fail to capture the enhanced retrapping of defects released directly adjacent to traps during detrapping or dissociation events.
This omission leads to systematic errors, including underestimated thermal desorption (TDS) peak temperatures and incorrect kinetic parameters when fitting to experimental data.
In this work, we derive for the first time analytical expressions for the \emph{adjacent sink strength}, including correction for finite impurity diffusion jump length.  
We provide a practical implementation strategy for integrating these expressions into kRE simulations.  
Comparisons with kinetic Monte Carlo (kMC) benchmarks demonstrate that adjacent sink strengths dominate the retrapping probability and are essential for
reproducing the correct temperature dependence of TDS release peaks.
Simulations that employ only random sink strengths can still be tuned to match TDS spectra; however, the resulting fitted trapping energies, 
detrapping frequencies, and diffusion parameters are often physically inconsistent.
The adjacent sink strength formulation introduced here significantly improves the predictive capability of kRE modeling, enabling accurate multiscale simulations of 
defect and impurity behavior in materials.
This framework also establishes a foundation for future extensions, including adjacent sink strengths associated with 
extended defects such as dislocations and grain boundaries, offering new opportunities to resolve persistent discrepancies between experimental and simulated trapping energetics.




\section{Introduction}

The physical and mechanical properties of materials depend critically on their chemical composition, defect and impurity concentrations, and internal microstructure.
Understanding and predicting material behaviour during ageing, ion irradiation, or thermal annealing therefore requires
multiscale modeling across a broad range of time and length scales.
Among the available approaches, the mean-field kinetic Rate Equations (kRE) method has become a widely used tool for simulating long-timescale defect evolution.
The method appears under several names, including cluster dynamics (CD), mean-field rate theory (MFRT), and master-equation approaches.
In all cases, the kRE framework treats defects as spatially varying concentrations (defects per volume), which
evolve in time under creation, annihilation, diffusion, trapping, detrapping, and reaction processes.
These processes form a set of coupled nonlinear differential equations that are solved in time and, in
spatially resolved formulations, in space \cite{McNabb63,Wiedersich72,Book_Freeman87,Stoller2008,Ahlgren12}.

The kRE methodology has been extensively applied to a wide variety of material phenomena.  
Examples include trapping and retention of helium (He) and hydrogen (H) isotopes in vacancies \cite{Baskes83,Myers86,Chroeun25},
irradiation‑induced amorphization and defect clustering \cite{SWAMINATHAN11,Katoh96,Rottler2005,Ortiz07,Mohamed_2025}, 
dislocation-loop and gas‑bubble evolution \cite{Wilson76,Wang17,Huang26}, 
precipitation, swelling, and segregation phenomena \cite{Wert50b,Brailsford72,Barnard12}, 
effect of edge-localized modes on fuel retention in tokamaks \cite{Heinola19},
hydrogen isotope exchange processes \cite{Schmid14,Hodille16,Markelj16}, 
and hydrogen isotope retention in advanced materials and next-step fusion devices \cite{Coenan17,Hodille26}.
The kRE method is also commonly used to simulate thermal desorption spectrometry (TDS) profiles \cite{Pisarev03,Poon08,Gasparyan15,GRIGOREV16}.

All physical parameters required for kRE applications must be obtained from experiments or lower‑scale simulations.
A key derived parameter is the \emph{sink strength}, which determines the rate at which mobile defects are trapped or annihilated.  
Sink strength depends primarily on trap size, concentration, and geometry \cite{Brailsford72};  
it also depends on the dimensionality of defect diffusion \cite{Trinkaus02}, diffusion jump length \cite{Hou16,Ahlgren17}, 
and, in some cases, elastic interactions between defects and traps \cite{Wolfer76,Rouchette14,Kohnert19}.
Analytical expressions for sink strengths exist for many symmetric trap geometries, including spherical traps, dislocation lines,
and grain boundaries \cite{Wiedersich72,Brailsford81}.  
For arbitrarily shaped traps, kinetic Monte Carlo (kMC) simulations are often used to determine sink strengths numerically \cite{Malerba07,Ahlgren17}.

Recent findings have shown that sink strength depends strongly on the \emph{initial position} of the defect \cite{Ahlgren20}.  
Traditional sink strengths describe defects created at random positions, which is appropriate for defects generated by ion irradiation.  
However, when a defect is created \emph{adjacent} to a trap, such as after detrapping or dissociation, the probability of retrapping increases dramatically.  
This enhanced retrapping corresponds to a much larger \emph{adjacent sink strength}.  
Neglecting this effect leads to substantial errors in kRE simulations, including incorrect TDS peak temperatures and temperature dependence, and misleading fitted parameter values, 
as demonstrated by comparisons between kRE and kMC simulations.

In this work, we derive for the first time analytical expressions for the adjacent sink strength, including correction for finite impurity jump length,
suitable for use in kRE and related mean‑field models.  
We demonstrate how these expressions should be applied, and we show that including the adjacent sink strength produces excellent agreement with kMC simulations across a range of conditions.  
For completeness, we also provide improved expressions for random-position sink strength.  
Finally, we discuss several caveats and inherent challenges associated with kRE modeling, and outline directions for future methodological advancements.

\section{Results}

The primary advantage of the kRE method is its computational efficiency, which often makes it the only feasible approach for simulating dynamic
defect processes in solids over long time and length scales. However, the reliability of kRE simulations depends critically on the accuracy of the applied sink strengths.

In this work, we distinguish between two types of sink strengths, depending on the initial defect position.
The {\bf random sink strength} describes the trapping probability for defects generated at random positions in the lattice.
The {\bf adjacent sink strength} applies to defects created immediately next to an empty trap, typically following a detrapping or dissociation event.

\subsection{Random position sink strength}

The analytical random-position sink strength for spherical traps in a system with 3D defect diffusion is given by the recursive expression (slightly modified from \cite{Brailsford81})
\beqn
K_R =  4\pi R_t C_t (1 + R_t \sqrt{K_{R,tot}}  ) \, \lambda_R^{Corr} / VF^{Corr}.    \label{eq:KR}
\eeqn
Here, $R_t$ is the trapping radius, $C_t$ the trap concentration, and $K_{R,tot}$ the total random sink strength. The factors 
$\lambda_R^{Corr}$ and $VF^{Corr}$ are the diffusion jump length ($\lambda$) and trap volume fraction corrections, respectively \cite{Ahlgren17}.
The jump length correction reduces the sink strength and is expressed as
\beqn
  \lambda_R^{Corr}=\mbox{exp}\bigg[- (1 + R_t\sqrt{K_{R,tot}})\big( P_{1,\lambda R_t}(\lambda/R_t) + P_{2,\lambda R_t} (\lambda/R_t)^2\big) \bigg],
\eeqn
where $P_{1,\lambda R_t}=0.295910$ and $P_{2,\lambda R_t}=0.050748$. This correction is applicable when $\lambda/R_t \lesssim 0.5$.
The trap volume fraction correction increases the sink strength and is given by
\beqn
  VF^{Corr}=\big[ 1 - P_{1,VF} VF^{P_{Exp,VF}} - (1 + R_t\sqrt{K_{R,tot}})P_{2,VF}(VF_{tot}-VF) \big],
\eeqn
with constants $P_{1,VF}=2.129798$, $P_{Exp,VF}=1.106332$ and $P_{2,VF}=1.165703$.
Here, $VF = C_t 4\pi R_t^3/3$ is the trap volume fraction and $VF_{tot}$ is the total volume fraction of all traps.
When only a single trap type is present, $(VF_{tot}-VF)=0$. The correction is valid for $VF_{tot} \lesssim 0.1$.

\subsection{Adjacent position sink strength}      \label{subsect:KA}

The analytical adjacent sink strength, derived in appendix \ref{app:KA}, is
\beqn
K_A&=&\frac{P}{1 - PD_t(2R_t+D_t)/6}   \lambda_A^{Corr}                 \qquad\qquad\qquad\qquad \ \ \     \mbox{: } k/\sqrt{C_{t,F}} \lesssim   0.2 \mbox{ nm}^{-1/2} \label{eq:KA_k0} \\
K_A&=&\frac{k^2[\alpha - (1-kL)]}{\alpha (\beta\mbox{e}^{-kD_t} - 1) - \ (1-kL)(\beta\mbox{e}^{kD_t} - 1)}\lambda_A^{Corr}\qquad \mbox{: }k/\sqrt{C_{t,F}} > 0.2 \mbox{ nm}^{-1/2}
\label{eq:KA_full}
\eeqn
where $k^2 = K_{R,E}$ is the random sink strength of empty traps of the same type with concentration $C_{t,E}$, and $C_{t,F}$ is the filled-trap concentration ($C_{t,E}+C_{t,F} = C_{t,tot}$).
The parameter $P = 4\pi R_tC_{t,F}(1+R_t/D_t)$, where $R_t$ is the trapping radius and $D_t$ the detrapping distance, and the remaining terms are\\
$\alpha = \mbox{e}^{-2k(L-R_t-D_t)}(1+kL)$, \ $\beta=(1+D_t/R_t)$ and \ $L=(3/(4\pi C_{t,F}))^{1/3}$.\\
The adjacent jump length correction factor is
\beqn
  \lambda_A^{Corr}=\mbox{exp}\big[- P_{\lambda R_t}(\lambda/R_t) - P_{1,\lambda D_t}(\lambda/D_t) + P_{2,\lambda D_t}(\lambda/D_t)^2 - P_{3,\lambda D_t}(\lambda/D_t)^3\big],
\eeqn
with constants: $P_{\lambda R_t}$= 0.374558, $P_{1,\lambda D_t}$=0.247795, $P_{2,\lambda D_t}$=0.010911, and $P_{3,\lambda D_t}$=1.860355\eex{-4} (App. \ref{app:KA}).
For $k \approx 0$, small $C_{t,F}$, and small $\lambda$, the adjacent sink strength reduces to
\beqn
  K_A \approx 4\pi R_tC_{t,F}(1+R_t/D_t),
\eeqn
which increases with the $R_t/D_t$ ratio. Physically, when a defect is detrapped very close to a large trap, the probability of retrapping increases substantially.
Note that if the detrapping distance is far from the trap ($D_t >> R_t$), the adjacent sink strength would conform to the classical low-concentration random sink strength $4\pi R_tC_t$.

\begin{figure}[h!]
  \centering
  \includegraphics[width=15.0cm]{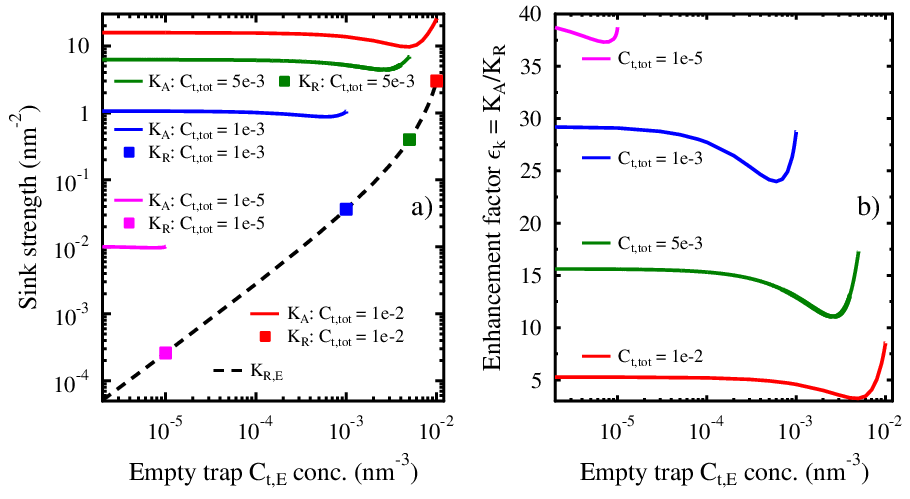}
  \caption{a) Random sink strength $K_R$ from Eq.~(\ref{eq:KR}) and adjacent sink strength $K_A$
    from Eqs.~(\ref{eq:KA_k0})-(\ref{eq:KA_full}) for varying trap concentrations ($C_{t,tot} = C_{t,E} + C_{t,F}$). 
    b) Corresponding sink strength enhancement factor. No jump length corrections were applied: $\lambda_A^{Corr}=\lambda_R^{Corr}=1$.}
  \label{fig:K_and_Eps}
\end{figure}

Figure~\ref{fig:K_and_Eps} shows the random and adjacent sink strengths along with the sink strength enhancement factor \cite{Ahlgren20} used in the kRE simulations.
The random sink strengths were iterated three times, Eq.~(\ref{eq:KR}). The trapping radius is 2 nm and the detrapping distance 0.05 nm.
The trap volume fraction for $C_{t,tot}=10^{-2}$ nm$^{-3}$ is about 0.1067.
The adjacent sink strength is approximately $4\pi R_tC_{t,F}(1+R_t/D_t)$ for $C_{t,F} > C_{t,E}$, decreases slightly when $C_{t,F} \approx C_{t,E}$,
and increases again when $C_{t,F} < C_{t,E}$.
The sink strength enhancement factor, Fig.~\ref{fig:K_and_Eps} b), is defined as the ratio of 
the sink strength for defects created adjacent to the trap divided by the sink strength for defects with random initial position $\epsilon_k = K_A(C_{t,F},C_{t,E})/K_R(C_{t,F}+C_{t,E})$.
The enhancement factor increases for smaller trap concentrations, however, its effect on the kRE simulations
remains crucial as the trap concentration increases, see Fig.~\ref{fig:OneTDScomp}.

\newpage

\subsection{Kinetic Rate Equation simulations}      \label{subsect:kREsim}

A limited presentation of the development of the sink strength and kinetic rate equation theories is given in
Refs. \cite{Wert50b,McNabb63,Wiedersich72,Brailsford81,Myers86,Ahlgren12,Ahlgren20}.
The time and 1D depth ($z$) evolution of a defect or impurity concentration, $I(z,t)$, can be related to diffusion, trapping, detrapping, and a source term
by the following equation \cite{Ahlgren20}
\beqn
   \frac{dI}{dt} &=& D\frac{d^2I}{dz^2} - D K_{R,tot} I + \frac{ C_{t,F} \nu_{detr}\ \mbox{exp}\big( -E_t/(k_{B}T)\big) } {\enhk} + S,   \label{eq:kRE}
\eeqn
where $D = \lambda^2\nu_{diff}\exp(-E_m/(k_BT))/6$ is the diffusion coefficient.
The second term after the equal sign describes trapping via the total random sink strength: $K_{R,tot} = \sum_i K_{R,i}$,
with each $K_{R,i}$ recursively evaluated from Eq.~(\ref{eq:KR}).
The third term represents detrapping reduced by retrapping \cite{Ahlgren20}, incorporating the sink strength enhancement factor.
$C_{t,F}$ is the filled trap concentration, $\nu_{detr}$ the detrapping attempt frequency, and $E_t$ the trapping energy.
The adjacent sink strength $K_A$ is determined from Eqs.~(\ref{eq:KA_k0}) or (\ref{eq:KA_full}) using the empty-trap random sink strength $K_{R,E}$,
while $K_R$ is computed using the sum trap concentration (empty + filled).

Next, we examine a few kRE simulations to highlight the effect that the adjacent sink strengths have on the results.
To validate the theory and equations, the results from kRE and kMC simulations will be compared.
The simulation conditions are chosen such that the corresponding kMC simulations can be carried out within a reasonable time, allowing a direct comparison.
All simulations are performed in a 100 nm thick material layer with one or three Gaussian trap profiles centered in the layer.
Initially, the traps are filled with one impurity per trap.
The temperature at the beginning of the simulation is set to 300 K, and it increases linearly at 50 K/s up to 800 K during the 10 s simulation.
As the temperature rises, detrapping and diffusion processes of impurities start to take place.
Some impurities are retrapped, but some reach the surface and leave the simulation layer.
Processes taking place on the surface have been omitted in the present simulations.
Thus, the impurity flux to the surface is taken directly as the flux of impurities crossing the front simulation layer boundary, 
which corresponds to standard Thermal Desorption Spectrometry (TDS), a commonly used experimental technique.

\begin{table*}[h!]
  \begin{minipage}{0.98\linewidth}
    \caption{Parameters used in the kMC and kRE TDS simulations with three Gaussian trap profiles (Fig.~\ref{fig:TDS3}). 
      $C_{t,max}$ (nm$^{-3}$) is the peak trap concentration of the Gaussian profile and SD (nm) its standard deviation. 
      Impurity diffusion parameters: jump length $\lambda = 0.05$ nm, frequency $\nu_{diff} = 2\eexeq{13}$ Hz, migration barrier $E_m=0.25$ eV. 
      Detrapping distance $D_t = 0.05$ nm.}
    \label{tab:threeTDSpars}
    \begin{tabular}{lllllll}
      \hline
      \hline
                         & Trap-Impurity pair  & \mc{2}{l}{Gaussian profile}    & \mc{3}{l}{ Trapping/detrapping parameters}  \\
                         &                     & $C_{t,max}$       & SD          & $R_t$  (nm) &  $E_t$ (eV) & $\nu_{detr}$ (Hz)\\
  \hline 
  Fig.~\ref{fig:TDS3} a) &   $C_{t,1}$-I        &   9\eex{-4}      &  7.0        & 1.0         &   0.95      & 5\eex{12}       \\
                         &   $C_{t,2}$-I        &   5\eex{-4}      &  9.0        & 1.2         &   1.15      & 2\eex{12}       \\
                         &   $C_{t,3}$-I        &   7\eex{-4}      &  8.0        & 1.5         &   1.35      & 3\eex{12}       \\
      \hline
      \mc{7}{l}{Fig.~\ref{fig:TDS3} b) Same parameters as in a) with added uniform impurity flux $2\times10^{-3}$ nm$^{-2}$s$^{-1}$} \\
      \hline
      \hline
    \end{tabular}
  \end{minipage}
\end{table*}
\begin{figure}[h!]
  \centering
  \includegraphics[width=15.0cm]{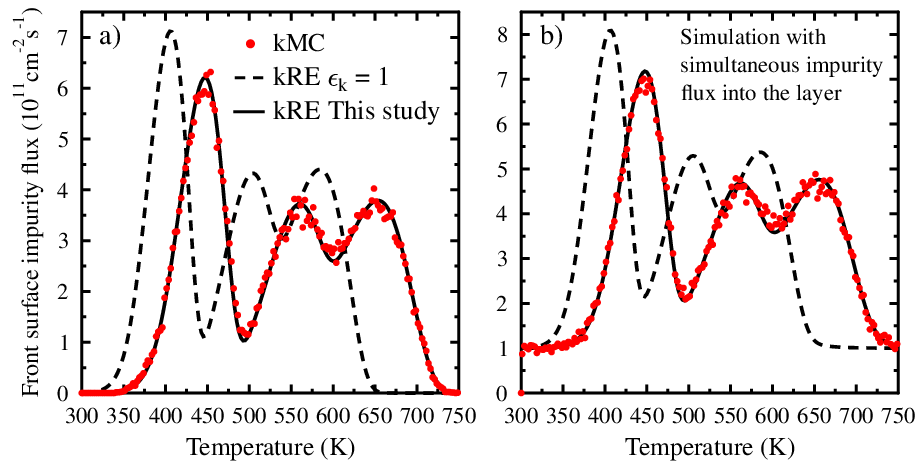}
  \caption{Front surface impurity flux from kMC and kRE TDS simulations for three trap profiles. Dash curves: kRE using only random sink strengths.  
    Solid curves: kRE using both random and adjacent sink strengths. a) no background flux; b) added uniform impurity flux $2\times 10^{-3}$ nm$^{-2}$s$^{-1}$.  
    Including adjacent sink strengths increases retrapping and shifts the TDS peaks to higher temperatures.}
  \label{fig:TDS3}
\end{figure}

Table~\ref{tab:threeTDSpars} shows the parameters, and Fig.~\ref{fig:TDS3} presents the results for the three Gaussian-shaped trap TDS simulations.
The kRE simulation scheme is elucidated in appendix \ref{app:Solv_kRE}.
The TDS peaks in Fig.~\ref{fig:TDS3} for the kRE simulations using only random sink strengths occur at much lower temperatures than they should.
The first TDS peak, corresponding to the lowest trapping energy 0.95 eV, appears about 40 K too early.
The TDS temperature-peak error increases with trapping energy, with the second peak appearing about 55 K, and the third about 70 K early.
Taking both random and adjacent sink strengths into account results in good agreement with the kMC simulations.
Fig.~\ref{fig:TDS3} b) shows the TDS results with an added simultaneous uniform impurity flux into the layer,
demonstrating that the theory ramains accurate even when additional random-position impurities are included.

\begin{figure}[h!]
  \centering
  \includegraphics[width=15.0cm]{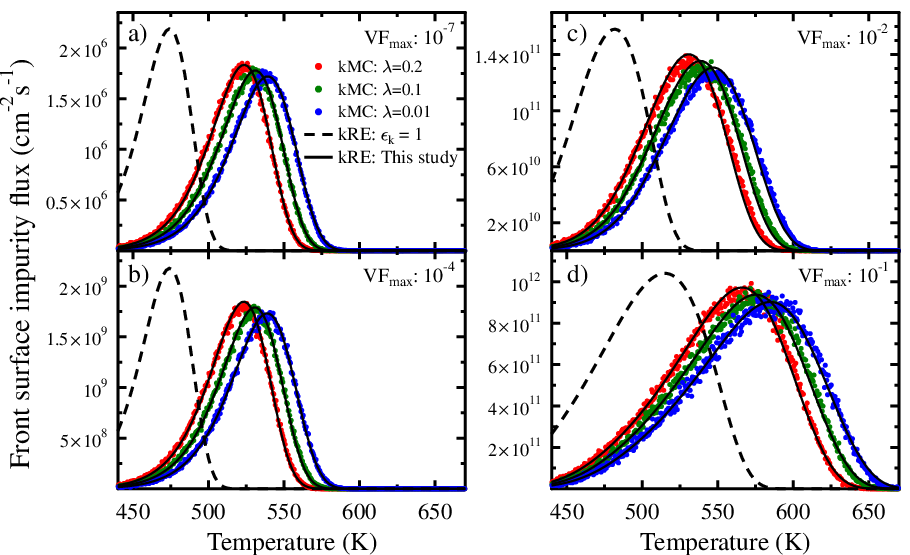}
  \caption{Front surface impurity flux from kMC and kRE TDS simulations for different peak trap volume fractions and diffusion jump lengths.  
    Higher trap concentrations shift the TDS peaks to higher temperatures.  
    Only kRE with adjacent sink strengths reproduces the trap concentration and jump length dependence observed in kMC.}
  \label{fig:OneTDScomp}
\end{figure}

We next examine the influence of trap concentration and diffusion jump length using a single Gaussian trap (SD = 5 nm).
The parameters were $R_t=2$ nm, $E_t=1.15$ eV, $\nu_{detr}=5\times10^{12}$ Hz, $D_t=0.05$ nm, and $E_m=0.25$ eV.
The peak trap volume fractions and corresponding concentrations in parentheses (nm$^{-3}$) were
\[
1\times10^{-7}\,(3\times10^{-9}),\;
1\times10^{-4}\,(3\times10^{-6}),\;
1\times10^{-2}\,(3\times10^{-4}),\;
1\times10^{-1}\,(3\times10^{-3}),
\]
with three jump lengths $\lambda$ = 0.2, 0.1, and 0.01 nm, and frequency $\nu_{diff}$ chosen such that the diffusion prefactor $D_0=1\times10^{10}$ nm$^2$/s remains constant.
Figure~\ref{fig:OneTDScomp} shows that the TDS peak maximum shifts to higher temperatures as the trap concentration increases.
This is mainly due to retrapping of detrapped impurities, where an impurity can either be retrapped in the adjacent trap from which it was detrapped or in a surrounding empty trap.
Both random and adjacent sink strengths increase as the trap concentration increases.
The increase in the random sink strength, and the corresponding TDS shift to higher temperatures, is visible in the dash curves in Fig.~\ref{fig:OneTDScomp}.
This TDS shift is approximately zero for very small trap concentrations, Fig.~\ref{fig:OneTDScomp} a)-b), where the random sink strength is very small.
It increases slowly to about 8 K for a peak trap volume fraction $VF_{max} = 10^{-2}$, and reaches about 40 K for the largest peak volume fraction $VF_{max} = 10^{-1}$.
The solid lines in Fig.~\ref{fig:OneTDScomp} show the results where both the adjacent and random sink strengths are taken into account.
We can see that the TDS shift due to retrapping to adjacent traps is already about 50 K at the lowest peak concentration
and remains large for all higher concentrations.
Besides the striking difference in the TDS peak positions due to trap concentration between kRE simulations using only random sink strengths
and those using both random and adjacent sink strengths, there is also a difference related to the impurity diffusion jump length.
Figure \ref{fig:OneTDScomp} shows that the kMC TDS peaks shift to higher temperatures as the diffusion jump length $\lambda$ decreases.
This effect is captured in the kRE simulations only when adjacent sink strengths are included,
while it is essentially absent when only random sink strengths are used (a tiny effect exists, but it is too small to be visible in the figure).
The jump length correction for random sink strength $\lambda_R^{Corr}$ is too weak to reproduce this effect, whereas the correction for adjacent sink strength depends on
both $\lambda/R_t$ and $\lambda/D_t$, leading to less retrapping for larger $\lambda$ and hence lower TDS peak temperature.

\section{Summary and discussion}

The number of studies employing multiscale kRE modeling has increased rapidly during the past decade. 
This growth is largely driven by the emergence of advanced kRE frameworks capable of treating complex reaction networks, 
making the method a practical bridge between atomistic simulations and experimental observations.
These complex reactions inevitably involve detrapping and dissociation processes
that demand a new class of sink strengths--\emph{adjacent sink strengths}--beyond the conventional random ones.

The goal of this work has been to provide analytical expressions and practical guidance for incorporating adjacent sink strengths into the kRE methodology.
We derived the adjacent sink strength formalism in detail, demonstrated how it should be implemented in simulations, and presented a
straightforward numerical approach for solving the coupled nonlinear equations.

As with any multiscale modeling approach, kRE and kMC simulations rely on parameters extracted from experiments or from smaller-scale methods such as quantum mechanics
(including electronic structure information) and molecular dynamics (including information at the level of individual atoms).
However, it is important to recognize that first-principles DFT energy barriers may have uncertainties of several tenths of an eV,
leading to rate-constant uncertainties of up to an order of magnitude.

\begin{figure}[h!]
  \centering
  \includegraphics[width=8.0cm]{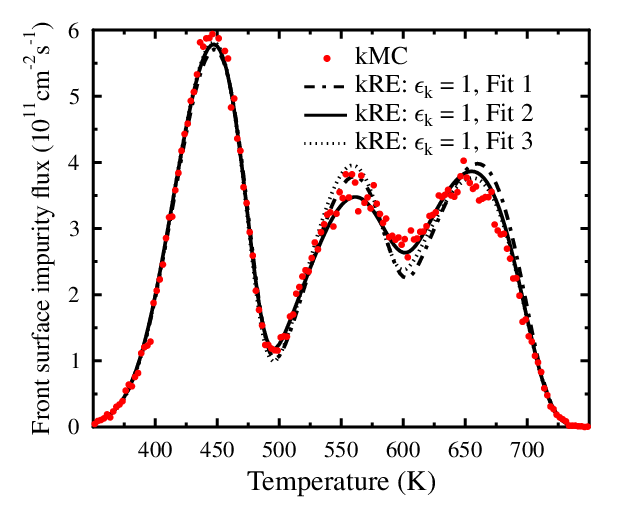}
  \caption{Different kRE fits to kMC TDS data (Table~\ref{tab:TDSdiscussion}). The kRE simulations neglect adjacent sink strengths ($\epsilon_k = 1$).  
    Despite good agreement with the kMC profiles, the fitted parameters are largely incorrect.}
  \label{fig:EpsOneFits}
\end{figure}

The kRE method itself introduces additional sources of error. The continuous differential equations are approximated on discrete spatial and temporal grids,
leading to discretization, truncation, and roundoff errors. Roundoff errors can be mitigated by using higher numerical precision, albeit at increased computational cost.
Discretization errors can usually be reduced by refining the grid, however, the resulting increase in the number of
numerical iterations may lead to cumulative errors in large-scale simulations.

kRE models are often used to extract parameter values by fitting simulations to experimental data.  
This approach requires caution, as illustrated by the additional kRE simulations performed for the three‑Gaussian‑trap TDS example.  
Figure~\ref{fig:EpsOneFits} shows that, when the adjacent sink strength is neglected,  
kRE simulations can still be tuned to match kMC TDS curves remarkably well, despite relying on incorrect parameter combinations.  
Table~\ref{tab:TDSdiscussion} demonstrates that the fitted diffusion and trapping parameters may deviate substantially  
from the true values used in the kMC simulation.  
Introducing additional fitting degrees of freedom (e.g., surface parameters) can further improve the visual fit,  
but typically pushes the fitted parameters even farther from their physically meaningful values.  

\begin{table*}[h!t!]
  \begin{minipage}{0.9\linewidth}
    {\small
    \caption{Fitting parameters for kRE TDS simulations using only random sink strengths (Fig.~\ref{fig:EpsOneFits}).  
      Bold values are fixed rather than fitted. The “Correct values’’ correspond to the parameters used in the kMC benchmark.}
    \label{tab:TDSdiscussion}
    \begin{tabular}{lllllllll}
      \hline
      \hline
                   & \mc{2}{l}{Diffusion} & \mc{6}{l}{ Trapping energies (eV) and detrapping frequencies (Hz) } \\
                   & $\nu_{diff}$ (Hz)     &  $E_m$ (eV) & $E_{t,1}$  & $\nu_{detr,1}$  & $E_{t,2}$  & $\nu_{detr,2}$ &  $E_{t,3}$ & $\nu_{detr,3}$\\
      \hline
   Correct  values & 2\eex{13}            & 0.25        &  0.95     & 5\eex{12}       &  1.15     & 2\eex{12}     &  1.35     & 3\eex{12}     \\
      \hline
   Fit 1           & 8.40\eex{13}         & 0.02        &  0.93     & 1.80\eex{11}    &  1.13     & 6.75\eex{10}  &  1.48     & 1.36\eex{12}  \\
   Fit 2           &{\bf 2x10$^{{\bf 13}}$}& {\bf 0.25}  &  0.92     & 1.42\eex{11}    &  1.27     & 1.29\eex{12}  &  1.59     & 9.13\eex{12}   \\
   Fit 3           & 3.06\eex{13}         & {\bf 0.35}  &  0.93     & 2.18\eex{11}    &  1.31     & 3.66\eex{12}  &  1.48     & 1.31\eex{12}   \\
      \hline
      \hline
    \end{tabular}
    }
  \end{minipage}
\end{table*}
This highlights a general principle:  
\emph{As the number of adjustable parameters increases, so does the number of non‑physical parameter combinations that can reproduce a given dataset.}  
Therefore, fits obtained using incomplete theoretical descriptions—such as omitting adjacent sink strengths—should be interpreted with great caution.

Keeping these limitations in mind, the kRE method remains a powerful, and, often the only feasible, approach for simulating defect and impurity evolution over
experimentally relevant time and length scales.
It provides valuable insight into experimental observations and can yield approximate parameter values when the relevant physics is correctly included.

Nevertheless, several open challenges remain. 
Increasing the complexity of reaction networks continues to push the demands on numerical solvers,  
necessitating ongoing refinement of integration schemes and discretization strategies.  
The intrinsic mean‑field nature of the kRE approximation could potentially be improved by incorporating  
local concentration gradients directly into the sink-strength formalism.  
Furthermore, new adjacent sink-strength models are needed for defects detrapped from extended microstructural sinks,  
such as dislocations and grain boundaries.  
Including such effects would shift simulated TDS peaks to higher temperatures  
and may help resolve discrepancies between experimentally measured and simulated trapping energies.

\section{Conclusions}

We have introduced a generalized sink-strength formalism for kRE simulations that explicitly accounts for 
defects created adjacent to traps following detrapping or dissociation events.  
This work derives analytical expressions for the adjacent sink strength, provides the necessary correction for realistic jump lengths,
and outlines a practical implementation strategy within the kRE framework.

By comparing kRE simulations with kMC benchmarks, we have demonstrated that the adjacent sink strength is essential 
for accurately describing retrapping processes.  
Simulations using only the conventional random sink strength systematically underestimate the TDS peak temperatures, 
whereas including the adjacent sink strength restores quantitative agreement with kMC over a wide range of trap concentrations, 
jump lengths, and trapping energies.  
These results show that adjacent sink strengths dominate the retrapping probability,
especially at low trap concentrations, where they exceed the random sink-strength contribution by more than an order of magnitude.

The study also reveals an important caution for parameter estimation:  
kRE simulations based on incomplete sink-strength formalisms can still be tuned to fit TDS spectra, 
but the resulting parameter sets are typically non-physical.  
This highlights the need for physically complete sink-strength models when using kRE simulations to infer trapping energies, 
detrapping frequencies, or diffusion parameters from experimental data.

Overall, the adjacent sink-strength formalism significantly improves the predictive capability of kRE modeling 
and enables more reliable multiscale simulations of defect and impurity evolution in solids.  
The methodology presented here establishes a foundation for further extensions, including adjacent sink strengths 
for defects detrapped from extended sinks such as dislocations and grain boundaries, 
which may help resolve persistent discrepancies between experimental and simulated trapping energetics.  
As multiscale models continue to increase in complexity, incorporating these refined sink-strength contributions 
will be essential for achieving accurate, long‑timescale predictions of microstructural evolution.

\pagebreak

\section{Appendix}

\subsection{Adjacent sink strengths}   \label{app:KA}

\begin{figure}[h!]
  \centering
  \includegraphics[width=7.0cm]{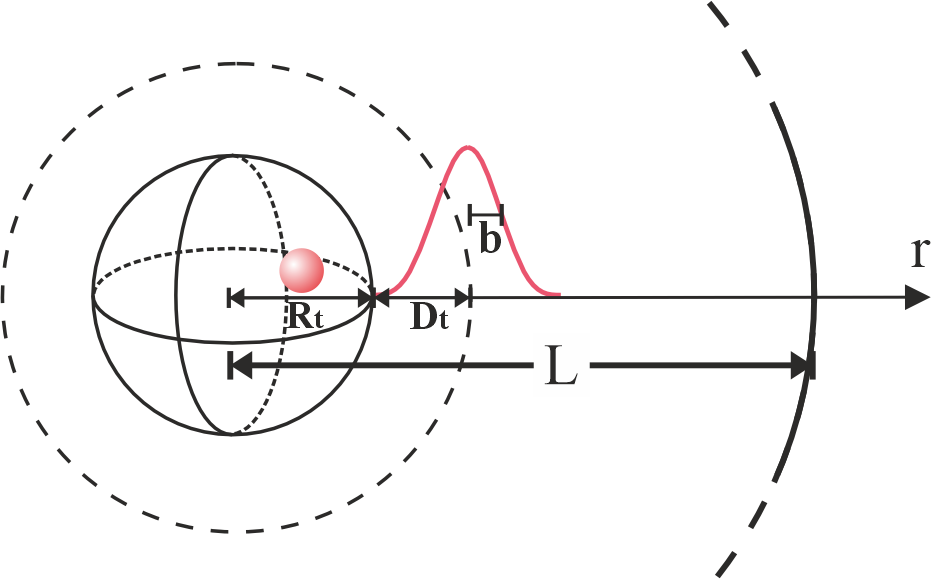}
  \caption{Schematic of the detrapping system. A defect detraps from a spherical trap (radius $R_t$) positioned at the center of a 
    spherical cell (radius $L$). The detrapping distribution is Gaussian with maximum at distance $D_t$ from the trap boundary and width $b$.}
  \label{fig:Detrapping_system}
\end{figure}

Figure~\ref{fig:Detrapping_system} illustrates the schematic geometry used to derive the adjacent sink strength. 
A defect is detrapped from a spherical trap of radius $R_t$, located at the center of a spherical cell of radius $L$.  
Detrapping is modeled as a Gaussian distribution with its maximum at a distance $D_t$ from the trap boundary and width $b$.

\begin{figure}[h!]
  \centering
  \includegraphics[width=9.0cm]{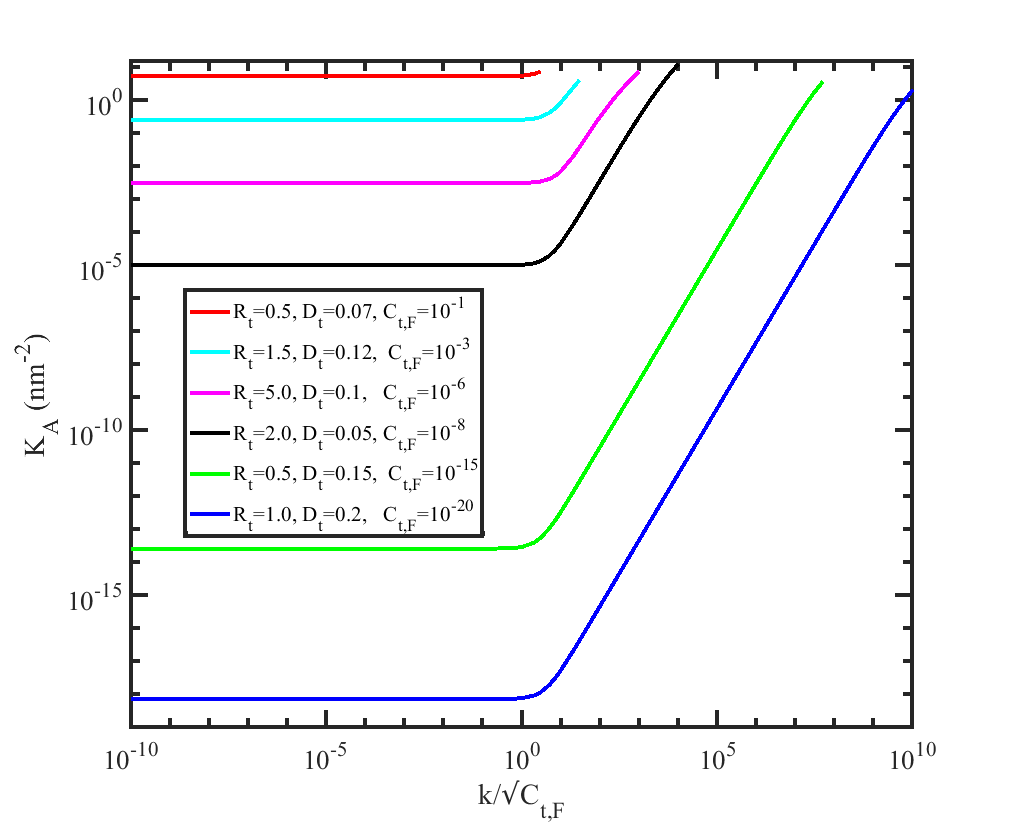}
  \caption{Adjacent sink strength as a function of $k/\sqrt{C_{t,F}}$ for different $R_t$ (nm), $D_t$ (nm), and $C_{t,F}$ (nm$^{-3}$) values.
    Eq. (\ref{Eq_KA_full}) is used for $k/\sqrt{C_{t,F}} > 0.2$, and Eq. (\ref{Eq_KA_k0}) for $k/\sqrt{C_{t,F}} \leq 0.2$.}
  \label{fig:KA_func_k}
\end{figure}

The time-dependent defect concentration $c(r,t)$ at distance $r$ from the trap center satisfies
\beqn
  \label{Eq_def_c_time}
  \frac{dc}{dt} = D \nabla^2 c - D k^2 c + \frac{A}{b\sqrt{2\pi}}\mbox{exp}\left[ -\frac{1}{2}\left( \frac{R_t+D_t-r}{b} \right)^2 \right],
\eeqn
where $D$ is the diffusion coefficient, $k^2 = K_R$ is the random sink strength of the surrounding traps, and the Gaussian source term 
represents detrapping with amplitude $A$ (other Gaussian parameters are defined in Fig.~\ref{fig:Detrapping_system}).
The cell boundary at $r=L$ is assumed to be equivalent to surrounding identical cells, giving a zero‑flux (Neumann) boundary condition.
The concentration in the cell increases with time and reaches a saturation distribution when the detrapping rate of
defects equals the sum of trapping to the central and surrounding traps in the cell.
This saturation distribution is given by the $dc/dt=0$ solution to the following equation in spherical coordinates, with parameter $A$ set to $D\sqrt{2\pi}$:
\beqn
  \label{Eq_def_c_diffeq}
  0 = \frac{d^2c}{dr^2} +\frac{2}{r}\frac{dc}{dr} - k^2c + \frac{1}{b}\mbox{exp}\left[ -\frac{1}{2}\left( \frac{R_t+D_t-r}{b} \right)^2 \right].
\eeqn
With boundary conditions $c(R_t) = 0$ and $\frac{dc}{dr}(L)=0$, we obtain a steady-state solution
{\small
\beqn
\label{Eq_def_c_diffeq_Sol1}
  c(r)&=&\frac{\sqrt{2\pi}}{rP_1} \Bigg(  \mbox{e}^{k(RD + r + kb^2/2)} \Bigg[  P_2\Big( \mbox{erf}\frac{D_t + kb^2}{\sqrt{2}b} -  \mbox{erf}\frac{RL + kb^2}{\sqrt{2}b} \Big)  \nonumber\\
       && + \mbox{e}^{-2kD_t} P_3\Big( \mbox{erf}\frac{Rr - kb^2}{\sqrt{2}b} -  \mbox{erf}\frac{D_t - kb^2}{\sqrt{2}b} \Big)  
          + \mbox{e}^{2k(L-RD)} P_4\Big( \mbox{erf}\frac{RL - kb^2}{\sqrt{2}b} -  \mbox{erf}\frac{Rr - kb^2}{\sqrt{2}b} \Big)  \Bigg]\nonumber\\
       && \qquad + \ \mbox{e}^{k(RD - r + kb^2/2)} \Bigg[ \mbox{e}^{2kL} P_5\Big( \mbox{erf}\frac{Rr + kb^2}{\sqrt{2}b} -  \mbox{erf}\frac{D_t + kb^2}{\sqrt{2}b} \Big) \nonumber\\
       && + \mbox{e}^{2k(L-D_t)} P_4\Big( \mbox{erf}\frac{D_t - kb^2}{\sqrt{2}b} -  \mbox{erf}\frac{RL - kb^2}{\sqrt{2}b} \Big) 
       + \mbox{e}^{2kR_t} P_2\Big( \mbox{erf}\frac{RL + kb^2}{\sqrt{2}b} -  \mbox{erf}\frac{Rr + kb^2}{\sqrt{2}b} \Big)  \Bigg]\nonumber\\
       &&\qquad - \frac{4kLb}{\sqrt{2\pi}} \mbox{e}^{\Big(-\frac{(L-RD)^2}{2b^2}+kL\Big)} \Big( \mbox{e}^{kr} - \mbox{e}^{k(2R_t-r)} \Big)  \Bigg),
\eeqn
}
with $L = \big(\frac{3}{4\pi C_{t,F}}\big)^{1/3}, \ RD = R_t + D_t, \ Rr = R_t + D_t - r, \ RL = R_t + D_t - L$,
and parameters $P_{1-5}$ as
\beqn
P_1 &=& 4k[(1 + kL)e^{2kR_t} - (1 - kL)e^{2kL}], \nonumber\\
P_2 &=& (1+kL)(R_t + D_t + kb^2), \qquad P_3 = (1+kL)(R_t + D_t - kb^2), \nonumber\\
P_4 &=& (1-kL)(R_t + D_t - kb^2), \qquad P_5 = (1-kL)(R_t + D_t + kb^2). \nonumber
\label{Eq_def_c_diffeq_Sol1_Pars}
\eeqn
When the detrapping width $b$ goes to zero, Eq.~(\ref{Eq_def_c_diffeq_Sol1}) simplifies substantially and splits into two regions relative to $r = R_t + D_t= RD$:
\beqn
\label{Eq_def_c_b0_1}
r \leq RD:&& c= P\Big( \mbox{e}^{kr} kLP \big[1 - \mbox{e}^{-2k(r-R_t)}\big]  +  \mbox{e}^{2k(L-D_t-r/2)}kLM \big[ 1 - \mbox{e}^{2k(r-R_t)} \big] \Big) \qquad \\
r >    RD:&& c= P\big(1 - \mbox{e}^{-2kD_t}\big) \big( \mbox{e}^{kr} kLP    -\ \mbox{e}^{2k(L-r/2)}    kLM  \big) \label{Eq_def_c_b0_3},
\eeqn
where $P = \frac{ \sqrt{2\pi}RD\mbox{e}^{k\cdot RD}   }{ 2kr[kLP \mbox{e}^{2kR_t} - kLM \mbox{e}^{2kL}] }$, $kLP=(1+kL)$, and $kLM=(1-kL)$.
From these simplified expressions, we extract the derivative at the trap boundary and mean defect concentration in the cell:
\beqn
\left.\frac{dc}{dr}\right|_{r=R_t} &=& Q/R_t   \Big[ \mbox{e}^{-2k(L-RD)}kLP - kLM \Big] \label{Eq_Der_Rt}\\
                        \meaneq{c} &=& \frac{3QR_t}{k^2L^3}  \Big[ \mbox{e}^{-2k(L-RD)} kLP (\beta\mbox{e}^{-kD_t} - 1) -  kLM (\beta\mbox{e}^{\ kD_t} - 1)  \Big], \label{Eq_MeanC}
\eeqn
where $Q = \frac{ \sqrt{2\pi}RD\mbox{e}^{k(2L - D_t)}   }{ kLP \mbox{e}^{2kR_t} - kLM \mbox{e}^{2kL} }$ and $\beta=(1+D_t/R_t)$.
The concentration of defects per unit time retrapped into the central trap (Fig.\ref{fig:Detrapping_system}) is proportional to
the trap concentration and area, and to the flux of defects at the trap boundary
$C_{t,F}\cdot 4 \pi R_t^2 \cdot D \left.\frac{dc}{dr}\right|_{r=R_t}$,
which in the mean-field approximation must equal $D\meaneq{c}K_A$, where $D$ is the diffusion coefficient.
Thus,
\beq
\label{Eq_KA_full}
K_A = \frac{k^2 [\alpha - (1-kL)]}{ \alpha [\beta\mbox{e}^{- kD_t} - 1]  - \  (1-kL) [\beta\mbox{e}^{\ kD_t} - 1] },
\eeq
where $\alpha = \mbox{e}^{-2k(L-R_t-D_t)}(1+kL)$.
The parameter $K_A$ is defined as the {\it adjacent} sink strength, i.e., the sink strength for defects formed adjacent to the trap
(in contrast to sink strength $K_R$ for defects formed in random positions).
The adjacent sink strength is proportional to the random sink strength for empty traps of same type, $K_{R,E}=k^2$.
For small $kL$ values (small random sink strength and/or large filled-trap concentration),
the denominator of Eq.~(\ref{Eq_KA_full}) approaches zero, and the limiting form becomes
\beq
\label{Eq_KA_k0}
K_A(k\rightarrow 0) = \frac{S}{1 - SD_t(2R_t+D_t)/6},
\eeq
where $S = 4\pi R_tC_{t,F}(1+R_t/D_t)$. This approximate solution is accurate when $k/\sqrt{C_{t,F}} \lesssim 0.2$ nm$^{-1/2}$,
as shown by the piecewise use of the two expressions in Fig.\ref{fig:KA_func_k}.

\begin{figure}[h!]
  \centering
  \includegraphics[width=12.0cm]{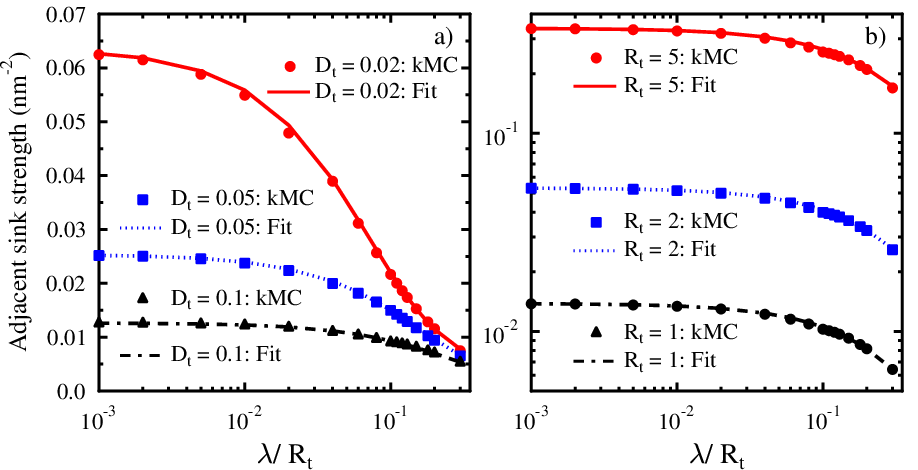}
  \caption{Analytical, Eq.~(\ref{eq:KA_k0}), and kMC adjacent sink strengths as a function of impurity jump length to trapping radius ratio.
    C$_{t,F}$ = 1\eex{-4} nm$^{-3}$.  a) R$_t$ = 1 nm.  b) D$_t$ = 0.1 nm. The  parameters for the fits are summarized in sub-section \ref{subsect:KA} in the jump length correction term.}
  \label{fig:KA_JLfits}
\end{figure}

Finally, kMC simulations using the {\it N}-jump method \cite{Ahlgren17} were performed to confirm the adjacent sink-strength equations
and their dependence on the impurity diffusion jump length.
The initial positions for the impurities are at a detrapping distance $D_t$ from the trap boundary \cite{Ahlgren20}.
Figure~\ref{fig:KA_JLfits} shows that $K_A$ decreases when either the detrapping distance or the diffusion jump length increases.
A similar jump-length decrease of the random sink strength has previously been observed and explained \cite{Hou16,Ahlgren17},
however, the decrease is much stronger and pronounced for the adjacent sink strength, see Fig.~\ref{fig:OneTDScomp} and related text.

\pagebreak

\subsection{Solving kinetic Rate Equations}   \label{app:Solv_kRE}

KRE simulations require solving large sets of coupled, nonlinear partial differential equations.  
Here, we summarize the solution method implemented in our HIM (Hydrogen Isotopes in Metals) program.

We use a finite‑difference (FDM) discretization with non‑uniform spatial grid points $i$ and time steps $j$.  
For a variable $I^{i,j}$, $i$ denotes depth and $j$ denotes time.  
The spatial grid spacing is refined where concentration gradients are large, and time steps are reduced where reaction rates  
(trapping, detrapping, implantation, etc.) are high.

Because the full system of equations is strongly coupled and nonlinear, a monolithic solution approach would be impractical.  
Instead, we use {\it Picard iteration} (successive substitution), solving one equation at a time using previously updated values of the other variables.  
Iterations continue until the solution converges within a predefined tolerance.

For the three‑Gaussian trap case (Sec.~\ref{subsect:kREsim}), the seven coupled equations are discretized in time using the second‑order 
Crank–Nicolson method \cite{fer81} and on spatially irregular grids following \cite{Vainonen01}.  
The discretized forms for the concentrations are:
\beqn
\frac{I\ijp1 - I^{i,j}}{\D t} &=& \frac{D}{2}\left[\frac{d^2I\ijp1}{dz^2} + \frac{d^2I^{i,j}}{dz^2} \right] + S\ijph - D K_{R,tot}\ijph [I\ijp1 + I^{i,j}]/2 \nonumber\\
      &&                                                    + \sum_{x=1}^3 \frac{ C_{F,x}\ijph \nu_{detr,x}\ \mbox{exp}( -E_{t,x}/(k_{B}T)) }{ {\enhk}_{,x}\ijph } \label{eq:I}\\
      \frac{C_{E,x}\ijp1 - C_{E,x}^{i,j}}{\D t} &=& - D K_{R,x}\ijph I\ijph + \frac{ C_{F,x}\ijph \nu_{detr,x}\ \mbox{exp}( -E_{t,x}/(k_{B}T)) }{{\enhk}_{,x}\ijph} \\   
      \frac{C_{F,x}\ijp1 - C_{F,x}^{i,j}}{\D t} &=& + D K_{R,x}\ijph I\ijph - \frac{ C_{F,x}\ijph \nu_{detr,x}\ \mbox{exp}( -E_{t,x}/(k_{B}T)) }{{\enhk}_{,x}\ijph},
\eeqn
with impurity $I$, and empty $C_{E,x}$ and filled $C_{F,x}$ traps of type $x$=[1-3]. The temperature $T$ and
impurity $I$ diffusion coefficient $D$ are evaluated in the middle of the time step $j+1/2$.
The second-derivative difference scheme for a non-uniform grid, where $\Delta z^i = z^i - z^{i-1}$, is
\beqn
 \frac{d^2I^i}{dz^2} &\approx&  \frac{ 2I^{i-1} }{ \D z^i (\D z^i + \D z^{i+1}) }
   -\frac{ 2I^i }{ \D z^i\D z^{i+1} }  +\frac{ 2I^{i+1} }{ \D z^{i+1}( \D z^i + \D z^{i+1} ) }.    \label{eq:secder}
\eeqn
$S\ijph$ is the implantation source term.
$K_{R,tot}\ijph$ is the total random sink strength of all empty traps, Eq.~(\ref{eq:KR}) iterated recursively three times, where $C_t$ is replaced by $C_{E,x}\ijph$.
$C_{F,x}\ijph$ is the filled trap concentration $x$, $\nu_{detr,x}$ is the detrapping frequency, and $E_{t,x}$ is the trapping energy.
${\enhk}_{,x}$ is the sink strength enhancement factor, see sub section \ref{subsect:kREsim} and Fig.~\ref{fig:K_and_Eps}.\\
\begin{enumerate}
    \item {\bf Solve} the tridiagonal Eq.~(\ref{eq:I}) for $I\ijp1$, compute $I\ijph$.
    \item {\bf Update} all $C_{E,x}$ and $C_{F,x}$ at time step $j+1$ and compute mid‑time values.
    \item {\bf Recalculate} the random sink strengths from the updated trap concentrations.
    \item {\bf Repeat} steps 1-3 until convergence is achieved (concentrations do not change more than some pre-determine value).
    \item {\bf Advance} to the next time step and repeat.
\end{enumerate}
The procedure continues until the final simulation time is reached.

\pagebreak

\section{References}

\bibliographystyle{unsrtnat}


\end{document}